\newcommand\apj{{ApJ}}% 
\newcommand\aap{{A\&A}}% 
\newcommand\nat{{Nature}}% 
\documentclass[dvips]{article}
\usepackage{icrctc07}
\title{\vspace*{-2.5ex}White Paper on the Status and Future of Ground-based Gamma-ray Astronomy
\vspace*{-1ex}
}
\shorttitle{Future Gamma-Ray Experiment}
\authors{
H.\,Krawczynski$^{1}$,\,J.\,Buckley$^{1}$,\,K.\,Byrum$^{2}$,\,C.\,Dermer$^{3}$,
\,B.\,Dingus$^{4}$,\,A.\,Falcone$^{5}$,\,P.\,Kaaret$^{6}$,
F.\,Krennrich$^{7}$,\,M.\,Pohl$^{7}$,\,V.\,Vassiliev$^{8}$,\,D.\,A.\,Williams$^{9}$\,for the white paper team.
}
\shortauthors{H. Krawczynski and et al.}
\afiliations{\footnotesize 
(1) Washington Univ. in St. Louis, 
(2) Argonne Nat.\ Lab.\, 
(3) U.S. Naval Research Lab., 
(4) Los Alamos Nat. Lab.,
(5) Penn State,
(6) Univ. of Iowa,
(7) Iowa State Univ.,
(8) UC Los Angeles,
(9) UC Santa Cruz \vspace*{-3.2ex}}
\abstract{In recent years, ground-based$\gamma$-ray observatories have made a number of important astrophysical discoveries which have attracted the attention of the wider scientific
community. The Division of Astrophysics of the American Physical Society has requested 
the preparation of a white paper on the status and future of ground-based $\gamma$-ray 
astronomy to define the science goals of the future observatory, to determine the 
performance specifications, and to identify the areas of necessary technology development. 
In this contribution we give a brief overview of the activities of the current 
white paper team and invite the international community to contribute to the
white paper.\vspace*{-2ex}}
\begin{document}
\maketitle
%Begin the section.
\section{Introduction}\vspace*{-2.5ex}
The field of TeV $\gamma$-ray astronomy was born in the years 1986 to 1988 
with the first firm detection of a cosmic source of TeV $\gamma$-rays
with the Whipple 10~m Cerenkov telescope, the Crab Nebula \cite{1989ApJ...342..379W}.
Advances in instrumentation and analysis techniques have established 
TeV $\gamma$-ray astronomy as one of the most exciting emerging new windows 
into the Universe. 

The current generation of ground based instruments includes imaging atmospheric
Cerenkov telescopes (IACTs) like H.E.S.S. \cite{2004NewAR..48..331H}, 
MAGIC \cite{2004NewAR..48..339L}, and VERITAS \cite{Meie:07} and water Cerenkov
arrays like MILAGRO \cite{Atkins:1999gb}. Arrays of IACTs achieve angular resolutions of 
0.15$^{\circ}$ and $\nu F_{\nu}$-sensitivities 
(250 GeV-1 TeV) of 10$^{-12}$ ergs cm$^{-2}$ s$^{-1}$ for 10 hrs 
of integration. Extensive air shower arrays have complementary capabilities to
IACTs. Whereas their instantaneous sensitivity is currently 
a factor of $\sim$150 lower than that of IACT experiments, their 
large field of view results in excellent survey sensitivity:
Milagro has surveyed 2$\pi$~sr of the sky at 20~TeV for point sources 
to a sensitivity of 2-4$\times$10$^{-12}$ ergs cm$^{-2}$ s$^{-1}$.
The past and current generations of experiments have proven that the TeV $\gamma$-ray sky is 
rich: more than two dozens of sources have been detected, including 
a wide range of galactic and extragalactic particle accelerators, e.g. 
the galactic center, super nova remnants, pulsar wind nebulae, 
X-ray binaries, and active galactic nuclei.
With an order of magnitude higher sensitivity, the next generation of ground 
based $\gamma$-ray experiments should be able to detect hundreds, maybe even thousands 
of sources. We are thus in the most exciting phase of an emerging field, 
when new source populations are still being discovered, and substantial 
samples of the strongest and most numerous sources can be studied.

Ground based $\gamma$-ray astronomy complements space-borne $\gamma$-ray astronomy
both, in terms of instrument capabilities and in terms of science questions that 
can be addressed. NASA and the Department of Energy will launch the
Gamma Ray Large Space Telescope (GLAST) in fall 2007 that will  
operate in the energy range from 30 MeV to 300 GeV.
Whereas GLAST will detect a large number of sources over a large field of view
and study flux variability on typical time scales from weeks to months, 
IACTs will detect fewer sources at higher energies with higher 
angular resolutions, and with better sensitivities on 
short (second to hours) time scales.

The American Physical Society (APS) has commissioned the writing of a white
paper which summarizes the science accomplishments of the field, and 
outlines the strategy to assure the continued progress of the field
on the long term. The APS charge included the formation of an editorial 
board  (Dingus, Halzen, Hofmann, Krawczynski, Pohl, Ritz, Vassiliev, Weekes) 
and the inclusion of input from all sectors of the physics, astrophysics, 
and astronomy communities.
The white paper\footnote{see http://cherenkov.physics.iastate.edu/wp/}
activities included the organization of several meetings
(e.g. a special session during the April 2007 APS meeting, and the meeting
''The future of ground based $\gamma$-ray astronomy'' in Chicago, May 2007) 
to widen the base of scientists involved in the discussion. The writing 
of the white paper will continue through 2007, and broad international 
participation is welcomed. In the following, we give a brief overview of the 
activities of six working groups that discuss the scientific and technical
achievements, challenges, and perspectives.
\vspace*{-2ex}
\section{Science Working Groups}
\vspace*{-2.5ex}
{\bf Supernova Remnants and Galactic Cosmic Rays:}
%
%The origin of cosmic rays and the mechanisms of their acceleration are
%among the most challenging problems in astroparticle physics and also
%among the oldest.  
Cosmic rays are energetically important in our
understanding of the interstellar medium (ISM) because they contain at
least as much energy as the other phases of the ISM.  Yet, the origin
of cosmic rays remains uncertain more than 90 years after their 
discovery by Victor Hess in 1912. 
Improving our knowledge of how high-energy particles are accelerated, 
diffuse and interact with the other components of the ISM in our Galaxy,
will help to understand other systems, such as active galactic
nuclei (AGN) that produce strong outflows with highly energetic particles.
Studies of particle acceleration in super nova remnants (SNRs) are of
particular interest, as their geometry is well constrained through 
observations at longer wavelengths. Furthermore, $\gamma$-ray images and
spatially resolved energy spectra can be obtained \cite{2006A&A...449..223A}.

The question of cosmic-ray acceleration in SNR includes aspects of the generation, 
interaction, and damping of turbulence in non-equilibrium plasmas. 
The physics of the coupled system of turbulence, energetic particles, and colliding plasma flows 
can best be studied in young SNRs. X-ray and TeV $\gamma$-ray observations 
indicate very efficient particle acceleration up to at least 100 TeV and the existence of a
turbulent magnetic field that is much stronger than a typical shock-compressed
interstellar magnetic field. The amplification of magnetic fields is of particular
interest because it may play an important role in the generation of cosmological magnetic fields. 
\\[1ex]
{\bf Galactic Compact Sources:}
Very high energy $\gamma$-ray emission is expected from 
Galactic sources such as pulsars both young and millisecond,
pulsar wind nebula, X-ray binaries containing neutron stars or
black holes, and colliding winds from massive stars.  The recent
H.E.S.S.\ Galactic plane survey revealed a rich variety of TeV emitters
\cite{2005Sci...307.1938A}.

Pulsar wind nebulae (PWNe) form when relativistic winds, powered
by the spin-down of young pulsars, terminate in a shock that
accelerates particles to energies reaching, perhaps, as high as
several PeV.  The prototypical TeV source, the Crab nebula, is a
pulsar wind nebula (PWN) and recent TeV observations have unveiled
many PWN, some previously unknown.  PWN offer a local laboratory
for the study of relativistic shock acceleration.  The also offer
a means to constrain the properties of pulsar winds and understand
the mechanism which dissipates the pulsar spin-down energy.  
New TeV observatories should detect a large set
($\sim 100$s) of PWN enabling population studies of 
how pulsar winds vary with spin-down power and age.
More sensitive TeV observatories will produce high fidelity,
energy-resolved maps of many PWN.  These maps, when combined with
maps at other wavelengths, will enable us to probe the physics of
particle acceleration in relativistic shocks and the diffusion of
relativistic particles.

Measurement of the cutoff energy of pulsed emission from
young and millisecond pulsar would provide a discriminant
between different models of the pulsar emission mechanism.
TeV light curves of young pulsars in binary systems would
allow us to extract information about the interaction of the
pulsar wind with the companion star outflow and enable direct
confrontation with magnetohydrodynamical simulations.  TeV
emission from black hole binary systems would provide a means to
determine the composition and total energy of the relativistic
jets produced by the black holes.
\\[1ex]
{\bf Extragalactic (non-GRB):}
TeV $\gamma$-ray observations of extragalactic objects afford the possibility
to study a wide range of phenomena. Active galactic nuclei (AGN) are spectacularly 
variable sources of TeV $\gamma$-rays.
More than a dozen BL Lac type AGN have now been identified as sources 
of $>$200~GeV $\gamma$-rays with redshifts ranging from 0.031 (Mrk 421) \cite{1992Natur.358..477P}
to 0.188 (1ES 0347-121) \cite{2006Natur.440.1018A}. The only extragalactic 
TeV $\gamma$-ray source that is not a BL Lac, is the radio galaxy 
M 87 \cite{2006Sci...314.1424A}. Future TeV $\gamma$-ray observations of AGN
hold the promise to reveal how supermassive black holes accrete matter and 
form powerful collimated outflows.

Owing to cosmic ray interactions with interstellar gas and subsequent 
$\pi_0$-decays, galaxies are expected to shine in $\gamma$-rays.
Indeed, the EGRET detector on board of the Compton Gamma Ray Observatory detected the 
Large Magellanic Cloud, located at the distance of $\sim 55$\,kpc.
Starburst and ultra-luminous galaxies with hard cosmic ray energy spectra 
should emit TeV $\gamma$-rays at a level close to the sensitivity of
current ground based experiments. An experiment with a one order of
magnitude higher sensitivity is expected to detect a considerable number 
of such galaxies, and will thus allow us to study the supernova/cosmic-ray 
connection in numerous extragalactic systems. Other potential extragalactic
sources of $\gamma$-rays include the largest particle accelerators in the Universe,
the lobes of radio galaxies, galaxy clusters, and large scale structure 
formation shocks.

TeV $\gamma$-ray observations have been used to set upper limits on the intensity 
of the extragalactic background light (EBL) in the optical/infrared wavelength
region. A next-generation experiment is likely to improve on these
results by a reliable detection of the absorption features that EBL photons
cause in the TeV $\gamma$-ray energy spectra of extragalactic sources owing
to pair-production processes. The measurement of the EBL intensity and
energy spectrum will make an important contribution to cosmology as the 
EBL depends on the universal structure and star formation history.

Other science topics that can be addressed with extragalactic $\gamma$-ray observations
include tests of Lorentz invariance and the measurement of intergalactic
magnetic fields. Although the chances for detecting the appropriate 
signatures are low, there is substantial discovery potential.
\\[1ex]
{\bf Gamma-Ray Bursts:}
Gamma ray bursts (GRBs) may emit both, prompt and afterglow 
GeV and TeV $\gamma$-ray emission. The dominant emission mechanisms 
and the opacity are almost certainly different for the prompt and the
afterglow phase. The $\gamma$-ray emission could
also come from the recently discovered late-time X-ray flares, which are
likely produced by late GRB internal engine activity and probably have
associated inverse-Compton emission. On theoretical grounds, both
short and long bursts could emit high-energy $\gamma$-rays.  The source
emission is predicted to be weaker for short bursts, but the attenuation
due to photon-photon interactions with the diffuse extragalactic
background will also be weaker for these bursts.
Measuring high-energy $\gamma$-ray emission from GRBs is of key importance 
for exploring the GRB environments and for constraining the efficiency of the
acceleration processes at work.  This information in turn 
contribute to identify the GRB progenitors. High-energy $\gamma$-ray 
observations have the potential to contribute to the identification 
of GRBs as ultra high energy cosmic ray accelerators.

Ground based $\gamma$-ray studies of GRB will require an experiment 
with significantly improved sensitivity, and either a large field of view
or fast-slewing narrow field of view instruments.
Furthermore, a lower energy threshold would increases the chances for positive
detections, as lower energy $\gamma$-rays suffer less extragalactic absorption.
\\[1ex]
{\bf Dark Matter:}
Another goal of the next generation $\gamma$-ray instrument will be to search for
$\gamma$-rays from dark matter annihilation in the halo of our own galaxy, or in
other galaxies.  In regions of enhanced halo density, weakly interacting dark
matter can annihilate to form a nearly mono-energetic $\gamma$-ray line as well as
a continuum of emission from annihilation through other channels (e.g.,
quark-antiquark, heavy leptons).  Any weakly interacting massive particle forms
a viable candidate for the dark matter.  The relic abundance of any particle in
equilibrium in the early universe is inversely proportional to the annihilation
cross-section and weakly interacting particles with masses $\sim$100 GeV could
provide densities close to the critical density.  The lightest supersymmetric
particle (neutralino) is the leading theoretical candidate, but any other
stable weakly interacting particle (e.g, the lightest Kaluza-Klein particle)
could also be a viable dark matter particle.  The possible mass for neutralinos
ranges from tens of GeV up to the unitarity limit around 100 TeV, but the
likely range of masses is 30 GeV to 3 TeV.  The signature of gamma-rays from
dark matter will be a mildly extended, cuspy angular distribution, a universal
continuum shape with a very hard spectrum and sharp cutoff and an annihilation
line at the mass of the neutralino. Outside of the galactic center, the best
places to look for dark matter are in galactic substructure or nearby Dwarf
galaxies.  In our own galaxy very nearyby microhalos (the first halo objects
formed in the early universe) could give an observable signal
\cite{2006PhRvL..97s1301K}.  Nearby intermediate mass black holes with halo spikes
could also be detectible \cite{2005PhRvL..95a1301Z}. 

Dark matter may be detected at the Large Hadron Collider or in
direct detection experiments, and neutrino experiments may provide a detection
of the dark matter in the local halo, gamma-ray measurements provide the only
possible means of observing the halo distribution and of verifying the role of
such particles in structure formation of the universe.
\\[1ex]
{\bf Technology:}
The baseline design of a next-generation instrument is determined by the requirement to 
achieve an one order of magnitude better sensitivity than the current instruments.
Furthermore, the science objectives call for increasing the energy bandwidth 
towards lower and higher energies, improving the angular resolution, and increasing the field of view
from between 3$^{\circ}$ and 5$^{\circ}$ diameter to between 6$^{\circ}$ and 12$^{\circ}$.
%The sensitivity goal for a next-generation instrument is illustrated in Figure 1. 
The sensitivity goal will require an experiment with a footprint-area on the order of 1 km$^2$.
The main component of the experiment would probably be an IACT array of mid-size (5\,m~-~15\,m diameter) 
telescopes. This main component could be complemented by a water Cerenkov array for 
large-field-of-view, high-duty-cycle observations.
Major design challenges are to reduce the cost per telescope and to minimize the operational costs.
Increasing the field of view may require to transition from Davies-Cotton or parabolic telescope
optics to Cassegrain optics. Other technology areas which could greatly impact sensitivity and 
cost include readout and trigger electronics design, the choice of photodetectors, and the
mirror fabrication technique.
\vspace*{-2ex}
\section{Outlook}
\vspace*{-2.5ex}
The previous and current generations of ground based $\gamma$-ray experiments have given us 
a first glimpse of the richness and uniqueness of the results that can be obtained at 
TeV energies. The white paper team has identified a large number of extremely interesting 
science topics in the fields of high-energy astrophysics, cosmology, and
particle physics that can be addressed by a next-generation TeV $\gamma$-ray experiment.
The team is still open for new team members, and for input 
from all sectors of the physics, astroparticle physics, and astronomy communities. For contact information, please see the white paper web-site$^1$.
The findings of the team will be published in Fall 2007.

The results obtained so far and the science potential clearly motivate the design and 
construction of a next-generation experiment which will detect hundreds or maybe thousands 
of sources. The development of such a next-generation experiment has started world-wide.
In the US, the next-generation experiment has been dubbed AGIS (Advanced Gamma-Ray Imaging System) 
and in Europe CTA (Cerenkov Telescope Array). Improving on the sensitivity of the current experiments 
by one order of magnitude will require substantial investments for R\&D and for the actual design, 
construction, and operation. It is clear that national and international collaboration will be instrumental. In the US and abroad, the next 
step will be R\&D over the next three to five years. 
Construction of the experiments could start in the years 2011 or 2012.
\vspace*{-2ex}
\section{Acknowledgments}
\vspace*{-2.5ex}
This work has benefitted from grant support from the U.S. Department of Energy, 
the U.S. National Science Foundation, and the Smithsonian Institution.\\[-5ex]
%\nocite{ref4}
%\nocite{ref5}
%\nocite{ref6}
%\nocite{ref7}


\begin{thebibliography}{10}

\bibitem{1989ApJ...342..379W}
T.~C. {Weekes} {\it et~al.}, \apj, {\bf 342},  379  (1989).

\bibitem{2004NewAR..48..331H}
J.~A. {Hinton}, NewAR, {\bf 48},  331  (2004).

\bibitem{2004NewAR..48..339L}
E. {Lorenz}, NewAR, {\bf 48},  339  (2004).

\bibitem{Meie:07}
G. {Meier et al.}, Procs. of the 30th ICRC, Merida: this volume~  (2007).

\bibitem{Atkins:1999gb}
R.~W. Atkins {\it et~al.}, NIM, {\bf A449},  478  (2000).

\bibitem{2006A&A...449..223A}
F. {Aharonian} {\it et~al.}, \aap, {\bf 449},  223  (2006).

\bibitem{2005Sci...307.1938A}
F. {Aharonian} {\it et~al.}, Science, {\bf 307},  1938  (2005).

\bibitem{1992Natur.358..477P}
M. {Punch} {\it et~al.}, \nat, {\bf 358},  477  (1992).

\bibitem{2006Natur.440.1018A}
F. {Aharonian} {\it et~al.}, \nat, {\bf 440},  1018  (2006).

\bibitem{2006Sci...314.1424A}
F. {Aharonian} and {et al.}, Science, {\bf 314},  1424  (2006).

\bibitem{2006PhRvL..97s1301K}
S.~M. {Koushiappas}, PhRvL, {\bf 97},  191301  (2006).

\bibitem{2005PhRvL..95a1301Z}
H. {Zhao} and J. {Silk}, PhRvL, {\bf 95},  011301  (2005).

\end{thebibliography}
\end{document}